%% file: main.tex
\documentclass[journal]{IEEEtran}

\usepackage{xcolor}
\usepackage{hyperref}
\usepackage{cleveref}

\usepackage{graphicx}
\usepackage{dcolumn}
\usepackage{bm}
\usepackage{epsfig}
\usepackage{float} 
\usepackage{amsmath}
\usepackage{amssymb}
\usepackage{subcaption}
\usepackage{tabularx}

\usepackage{stfloats}
\usepackage{comment}
\usepackage{longtable}
\usepackage{url}

\usepackage{cite}

\extrafloats{100}
\input{defs}

\begin{document}


\title{A Comparison of the Effects of Neutron and Gamma Radiation in Silicon Photomultipliers}

\author{
 B.~Bir\'o,
 G.~David,
 A.~Fenyvesi,
 J.S.~Haggerty,
 J.Kierstead,
 E.J.~Mannel,
 T.~Majoros,
 J. Molnar,
 F.~Nagy,
 S.~Stoll, 
 B.~Ujvari,  
 C.L.~Woody
  \thanks{Please see Acknowledgements for author affiliations.}
}

\maketitle

\begin{abstract}

The effects of radiation damage in silicon photomultipliers (SiPMs) from gamma rays have been measured and compared with the damage produced by neutrons. Several types of MPPCs from Hamamatsu were exposed to gamma rays and neutrons at the Solid State Gamma Ray Irradiation Facility (SSGRIF) at Brookhaven National Lab and the  Institute for Nuclear Research (Atomki) in Debrecen, Hungary. The gamma ray exposures ranged from 1 krad to 1 Mrad and the neutron exposures ranged from 10$^8$ n/cm$^2$ to 10$^{12}$ n/cm$^2$. The main effect of gamma ray damage is an increase in the noise and leakage current in the irradiated devices, similar to what is seen from neutron damage, but the level of damage is considerably less at comparable high levels of exposure. In addition, the damage from gamma rays saturates after a few hundred krad, while the damage from neutrons shows no sign of saturation, suggestive of different damage mechanisms in the two cases. The change in optical absorption in the window material of the SiPMs due to radiation was also measured. This study was carried out in order to evaluate the use of SiPMs for particle physics applications with moderate levels of radiation exposures.

\end{abstract}

\begin{IEEEkeywords}
SiPM, MPPC, Radiation Damage, Gamma Rays, Neutrons, sPHENIX, EIC, Calorimeters
\end{IEEEkeywords}



\section{Introduction}
\input{introduction}

Numerous studies have been carried out on the effects of radiation damage in silicon photomultipliers (SiPMs) due to neutrons and heavy charged 
particles~\cite{Matsumura:2009he,NAKAMURA2009110, Bohn:2008gc,MUSIENKO200987,Angelone:2010mg,Qiang:2012zh,Andreotti:2014uma,
Musienko:2015lia,LI201663,Heering:2016lmu,Tsang:2016cmc,Cattaneo:2017vgl, Musienko:2017znn,CentisVignali:2017,Garutti:2017ipx}. These studies have all shown that neutrons with energies $\sim$ 1 MeV cause a significant increase in noise and leakage current that can lead to difficulties in using these devices in applications with high levels of neutron fluence. Additional studies have been carried out with thermal 
neutrons~\cite{Durini:2016uzu} which have shown similar effects, but only at very high fluences. There have been fewer studies on the effects of gamma ray irradiation on 
SiPMs~\cite{Garutti:2014jya,Pagano:2014bua,Xu:2014vua}, and those studies show a similar effect of increased leakage current with dose, but typically to a much lesser degree. However, gamma rays can also cause other forms of damage in certain SiPMs, such as producing optical absorption in the protective window of the device (which is typically epoxy), that can lead to a loss in photon detection efficiency. 

    It is well known that neutrons, as well as protons, with energies $\sim$ 1 MeV cause significant damage in silicon due to the production of defects in the bulk material. This occurs when a neutron or proton with sufficient energy knocks a silicon atom out of its normal position in the lattice to an interstitial site, leaving a vacancy. This mechanism, known as displacement damage, or the formation of Frenkel defects, has been extensively studied and documented in the literature~\cite{Gill:1997dt,Srour:2003,Srour:2013,Moll:2018}. These defects can be either single point defects or cluster defects that can extend over distances of several hundred Angstroms in the lattice. The formation of these defects produces energy levels in the band gap, in addition to those that are already present due to intrinsic defects, making it easier for electrons in the valence band to be promoted to the conduction band due to thermal excitation, thus increasing the intrinsic noise in the device. Depending on the nature of these radiation induced defects, they can potentially be removed by thermal annealing, where the degree of recovery is dependent on the temperature and time after exposure. Defects produced by high energy neutrons or protons which cause nuclear breakups in the lattice cannot be recovered.   
    
    Low energy neutrons can also cause damage in silicon that can lead to similar effects. Thermal neutrons can be captured by a silicon atom leading to nuclear transmutations, such as $^{30}$Si + n $\rightarrow$ $^{31}$Si $\rightarrow$ $^{31}$P + $\beta$. These defects are also permanent and cannot be recovered, but the probability for their formation depends on the thermal neutron capture cross section in silicon, which can be much lower than the cross section for higher energy neutron interactions. Thermal neutrons can also be captured on various dopant materials implanted in the device, such as boron, but the degree of capture depends on the dopant concentration which is dependent on the manufacturing process. 
    
     Gamma rays, as well as electrons or positrons, can produce damage in SiPMs by several mechanisms. High energy gammas or electrons can also produce displacement damage, but the cross section for a large momentum transfer to a silicon nucleus is smaller than for heavier charged or neutral particles. The damage mechanism for gamma rays and electrons is also discussed in the literature \cite{Summers:1993,Xapsos:1994,ElAllam:2018}. These particles tend to produce more single point defects and are less effective in producing displacement damage than neutrons or heavier charged particles. However, gammas and electrons can also produce ionization damage, leading to charging up effects inside the device that can distort the electric fields, and can also produce absorption in the entrance window due to the formation of optical absorption bands. Finally, even low energy X-rays can cause surface damage to the SiO$_2$ layer of the device which can affect its performance.
     
    The main purpose of this study was to investigate the use of silicon photomultipliers for use in particle physics applications with moderate levels of radiation exposures, such as the sPHENIX experiment at RHIC \cite{sPHENIX:2016TestBeam} where they will be exposed to neutron fluences on the order of $10^{10}$ n/cm$^2$ per year and ionizing radiation doses of a few tens of krad per year, or at the future Electron Ion Collider being proposed at Brookhaven National Lab or Jefferson Lab. The devices in this study were irradiated with  neutrons up to total integrated fluences of 10$^{12}$ n/cm$^2$ and total gamma ray doses up to 1 Mrad. Several types of Hamamatsu MPPCs were tested in order to study how devices with different pixel sizes would perform in such a radiation environment and the effects after long term exposure. 

\section{Experimental Methods}

Neutron and gamma ray irradiations were carried out at two different facilities: the Solid State Gamm-Ray Irradiation Facility (SSGRIF) at Brookhaven National Lab and the cyclotron based fast neutron source at the Institute for Nuclear Physics Research (Atomki) in Debrecen, Hungary. By carrying out irradiations at two different facilities, it provided a way to estimate the systematic uncertainties between the two sets of measurements. 

  The SSGRIF at BNL consists of a large $^{60}$Co source that is used for gamma ray irradiations which is capable of producing dose rates between 0.1 and 50 krad/hr. The exposures for this study were done at 10 krad/hr. The total integrated gamma ray doses were determined with a precision of 5\% using Optically Stimulated Luminescence (OSL) dosimetry. Neutron exposures at the SSGRIF utilized a D-T generator to produce $\sim$ 14 MeV neutrons with a flux $\sim$ 10$^5$ n/cm$^2$/sec, which limited the total exposures to a few times 10$^{10}$ n/cm$^2$. 
  
  \begin{figure}[h]
 \begin{center}
	\includegraphics[width=9cm]{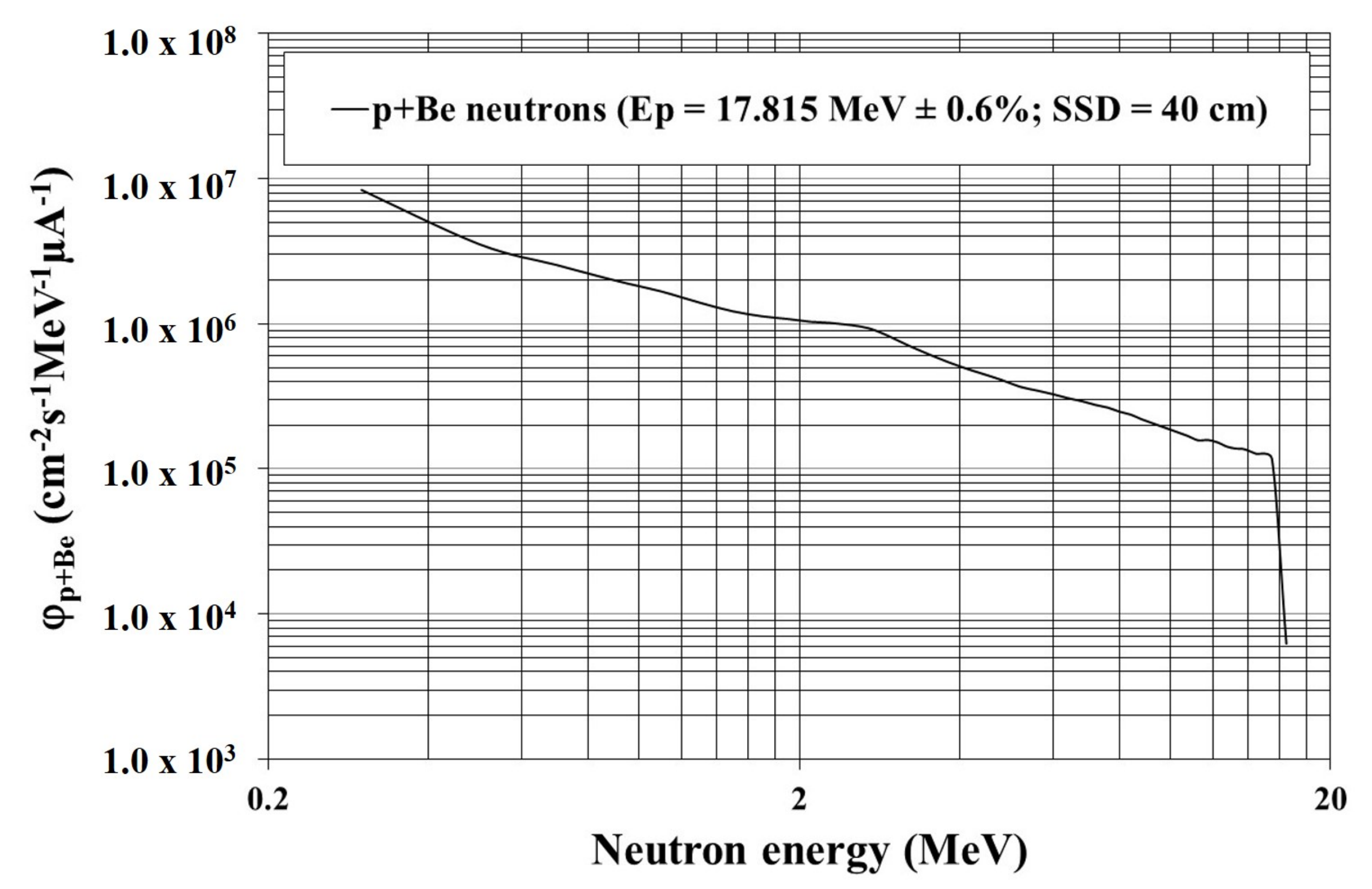}
   	\caption{\label{fig:Atomki_neutron_spectrum} Flux of neutrons as a function of energy per unit beam current produced at the irradiation position at the $p$+Be neutron source at Atomki.  
    }
 \end{center}
\end{figure}
  
  Neutron exposures at Atomki utilized a cyclotron to bombard a 3\,mm thick beryllium target with protons with an energy of $E_p$=17.815\,MeV ($\pm$0.6\%). This produced a spectrum of neutrons with energies up to $\sim$ 17 MeV and allowed achieving much higher neutron fluences. Figure~\ref{fig:Atomki_neutron_spectrum} shows the neutron spectrum per unit beam current at the irradiation position interpolated from the data by Brede {\it et al.}~\cite{Brede:1989}. The neutron flux could be changed by adjusting the beam current to be in the range from 1.5 $\times 10^{5}$ - 8.0 $\times 10^{7}$ n/cm$^2$/sec. The absolute neutron flux was determined by measuring the beam current on the target, as well as using pair of twin ionization chambers that also measured the accompanying gamma ray dose using the method described in ~\cite{Broerse:1981}. It is estimated that the absorbed gamma ray dose for these measurements was $\sim$ 5 Gy and that the overall uncertainty the absolute neutron fluence was $\pm$ 15\%.  

  The total flux of neutrons impinging on the sample $\Phi_{p+Be}$ was converted to an equivalent flux of 1 MeV  neutrons $\Phi_{E_n=1MeV}$ using the Non-Ionizing Energy Loss (NIEL) scale factor $\kappa_{p+Be}(Si)$ for neutrons on silicon according to the formula: 
  
\begin{equation}\label{eq:1MeV_neutron_equiv_p+Be}
\Phi_{E_n=1MeV} = \kappa_{p+Be}(Si)*\Phi_{p+Be}
\end{equation}

  The factor $\kappa_{p+Be}(Si)$ was computed using the  atomic displacement damage cross section for silicon D($E_n$) from Griffin \cite{Griffin:1993} integrated over the neutron spectrum shown in Fig.~ \ref{fig:Atomki_neutron_spectrum} normalized to the damage cross section for 1 MeV neutrons D($E_n = 1 MeV$), as given by the following formula:

\begin{equation}\label{eq:kappa_p+Be}
\resizebox{0.9\hsize}{!}{$ 
\kappa_{p+Be} = \frac{1}{ D(E_n=1MeV) } \times \frac{ 
{\int_{0}^{E_{n,max}} D(E_n)*\Phi_{p+Be}(E_n)dE_n} }
{ {\int_{0}^{E_{n,max}}}   \Phi_{p+Be}(E_n)dE_n}
$}
\end{equation}

  Neutrons from the D-T source at the SSGRIF are essentially monoenergetic with an energy of 14 MeV. The NIEL scale factor used to convert the flux of these neutrons to an equivalent flux of 1 MeV neutrons was 1.787 and was also taken from Griffin \cite{Griffin:1993}. However, the true flux could only be estimated from the nominal flux for the D-T generator given by the manufacturer and therefore had a much larger uncertainty compared to the flux measured at Atomki. 

   All of the samples tested in this study were Hamamatsu 3 mm x 3 mm Multipixel Photon Counters (MPPCs). Three types of devices were irradiated in order to study the dependence of the damage on the pixel size. These were Hamamatsu S12572-015P with 15 $\mu$m pixels, S12572-025P with 25 $\mu$m pixels and S13360-3050PE with 50 $\mu$m pixels. The S13360s were produced using an improved technology that reduced their dark current, cross talk and after pulsing compared to the older technology used to produce the S12572s. Hamamatsu now produces 3 x 3 mm$^2$ devices with smaller pixel sizes using this improved technology which we plan to investigate in a future study. 

  The currents for all the devices were measured using high precision picoammeters (Keithley 6487 with 10 fA resolution at BNL and Keithley 2635B with 0.1 fA resolution at Atomki). For the sequence of gamma ray and neutron irradiations, the devices were irradiated at their normal operating bias and the currents were measured at the manufacturer's recommended operating voltage which varied from device to device. All measurements were done at nominal room temperature (typically 23-25 $^{\circ}$C). During irradiation the temperature was generally controlled to $\sim$ 1-2 $^{\circ}$C, while measurements done in the laboratory were done in a temperature controlled box where the temperature was held constant to $\sim$ 0.5 $^{\circ}$C.

\section{Results}

\subsection{Gamma ray irradiations}

\begin{figure}[th]
 \begin{center}
   \includegraphics[width=9cm]{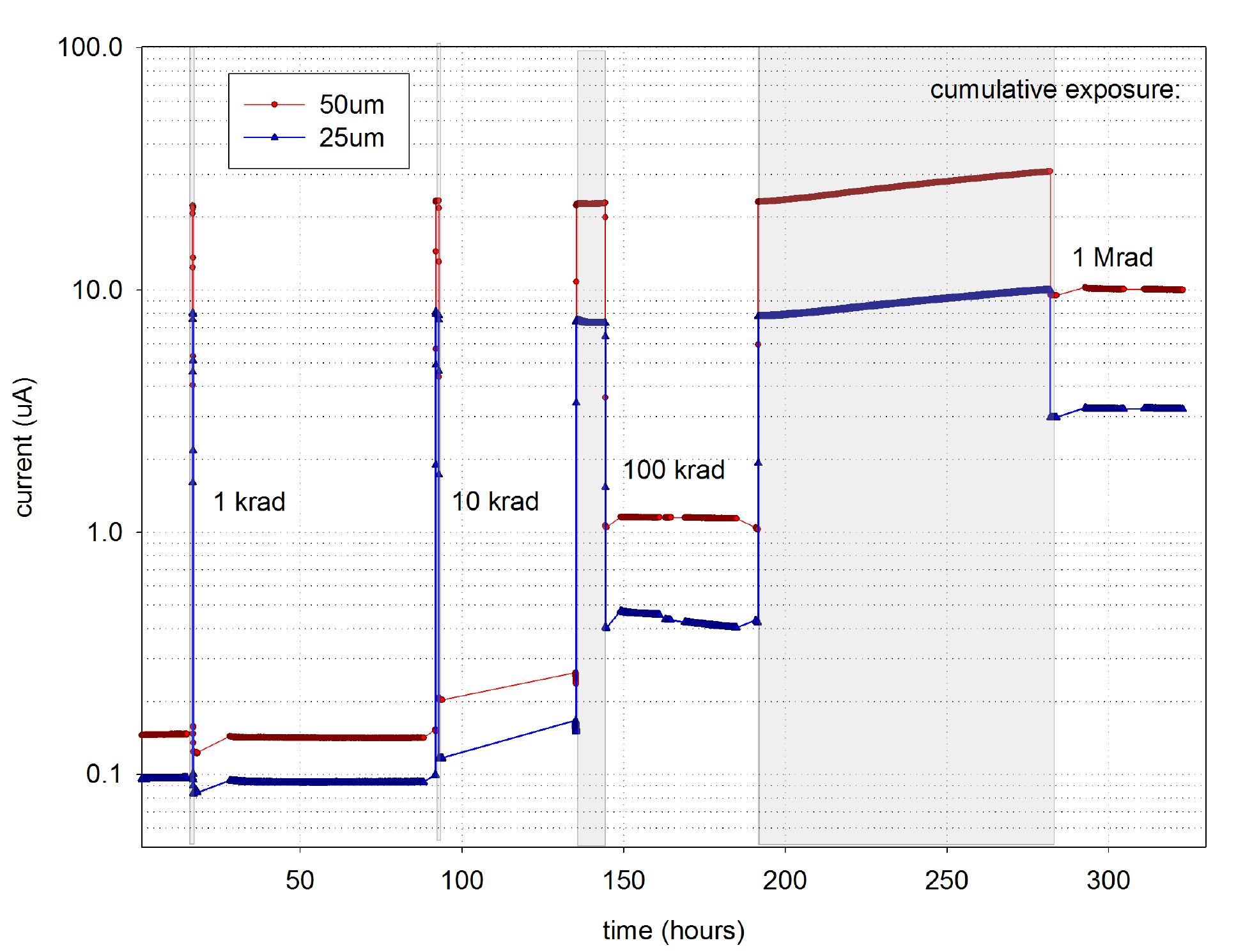}
   	\caption{\label{fig:SSGRIF_gamma_ray_irradiation} Sequence of gamma ray irradiations of Hamamatsu S12572-025P and S13360-3050PE MPPCs from 1 krad to 1 Mrad with the $^{60}$Co source at the BNL Solid State Gamma-Ray Irradiation Facility (SSGRIF). Currents were measured at the manufacturers recommended operating voltage for each device as listed in Table~\ref{tab:radiation_induced_currents}.  }
 \end{center}
\end{figure}

  Gamma ray irradiations at the SSGRIF were carried out in a series of steps of increasing dose, starting with an integrated dose of 1 krad and increasing in steps by one order of magnitude up to 1 Mrad. Time was taken between the exposures to measure the samples and to observe any change in current over time due to annealing. Figure~\ref{fig:SSGRIF_gamma_ray_irradiation} shows a sequence of gamma ray irradiations for 25 $\mu$m and 50 $\mu$m pixel devices. The current increases dramatically during irradiation due to the large current induced by gammas from the source, but this induced current drops sharply when the source is removed. 
   For low doses, the current drops immediately after exposure to essentially its initial value. Then, as the dose increases, the baseline current increases and shows very little effect of annealing. For a cumulative dose of 1 Mrad, the dark current for the 25 $\mu$m device increased to 3 $\mu$A, while the dark current for the 50 $\mu$m device increased to 10 $\mu$A.  

\begin{figure}[th]
 \begin{center}
   \includegraphics[width=9cm]{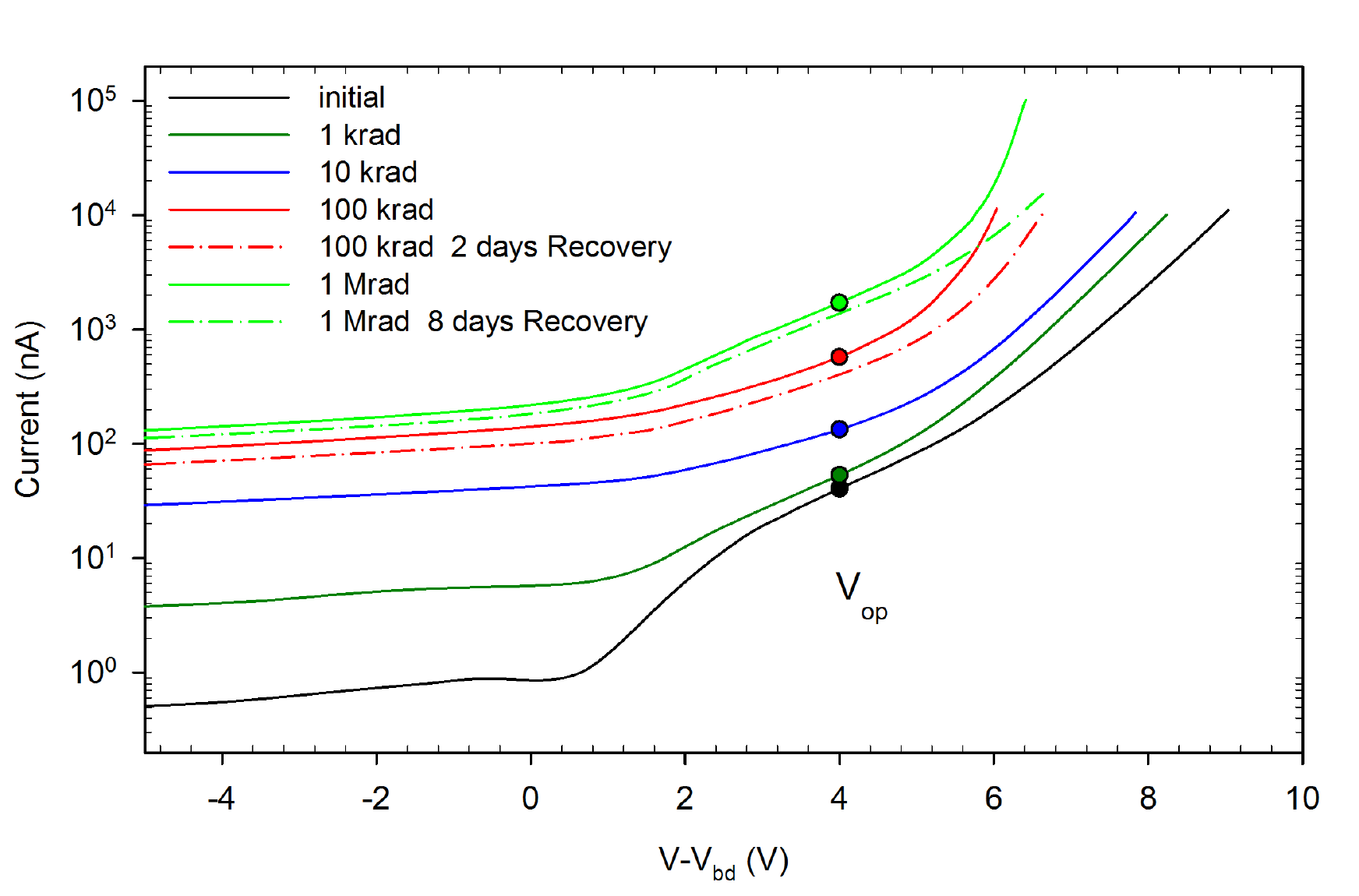}
   	\caption{\label{fig:1Mrad_gamma_irradiation_and_recovery} Gamma ray irradiation and room temperature recovery of a Hamamatsu S12572-015P 15 $\mu$m pixel device at the BNL SSGRIF.}
 \end{center}
\end{figure}

Figure~\ref{fig:1Mrad_gamma_irradiation_and_recovery} shows the effect of damage and recovery for a 15 $\mu$m device (Hamamatsu S12572-015P) for various steps of increasing dose up to 1 Mrad. The results show a large increase in dark current both above and below the breakdown voltage and the amount of increase is similar in the two regions. This indicative of surface damage which can lead to a large increase in leakage current outside the avalanche region and has been observed in other measurements \cite{Xu:2014vua}. The recovery curves are for periods of 2 and 8 days at room temperature. There is virtually no recovery for doses up to 10 krad, and only a small recovery for doses of 100 krad and 1 Mrad which appears to be rather uniform across the I-V curve.

\begin{figure}[th]
 \begin{center}
   \includegraphics[width=9cm]{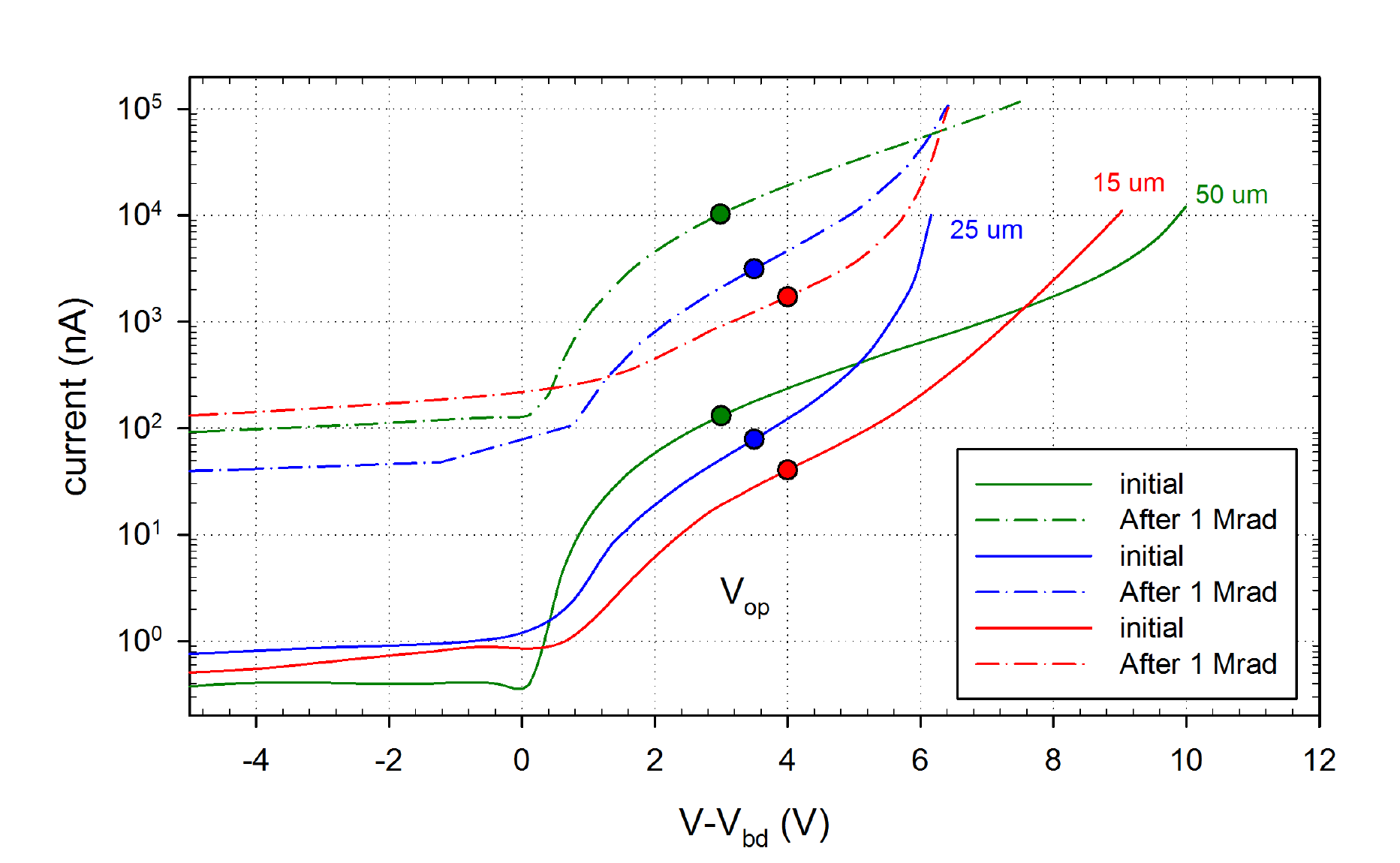}
  	\caption{\label{fig:Gamma_irradiation_3_I-V_curve} I-V curves for 15 $\mu$m, 25 $\mu$m and 50 $\mu$m pixel devices irradiated with gamma rays up to 1 Mrad at a dose rate of 10 krad/hr at the BNL SSGRIF.}
 \end{center}
\end{figure}

Figure~\ref{fig:Gamma_irradiation_3_I-V_curve} shows the I-V curves for devices with three different pixel sizes (Hamamatsu 12572-015P, -025P and S13360-3050P) irradiated up to a dose of 1 Mrad at the BNL SSGRIF. All devices show a large increase in dark current, both above and below their breakdown voltages, similar to what was observed in Figure \ref{fig:1Mrad_gamma_irradiation_and_recovery}. 

  In order to compare the three different devices, we have computed the radiation induced currents divided by their active area (correcting for the different fill factors) and divided by the gain of each device at its nominal operating voltage. The operating voltage V$_{op}$ is defined as the voltage above the breakdown voltage (V$_{bd}$+V) that is required to achieve the nominal gain specified by the manufacturer for a given device. These normalized currents are given in Table~\ref{tab:radiation_induced_currents} for their initial values and after a dose of 1 Mrad. All the devices reach a similar level of induced current $\sim$  1.0-1.5$\times 10^{-3}$ nA/G/mm$^2$ after exposure. The S13360 showed about a factor of two higher increase relative to its initial current, which may be partly due to the lower initial current for these devices that were produced with the newer technology.   

\begin{table}[!hbt]
\centering


\caption{}

\center{RADIATION INDUCED CURRENTS FROM GAMMA RAYS AND NEUTRONS IN DIFFERENT PIXEL SIZE DEVICES}
\vspace{0.1in}

\resizebox{1.0\hsize}{!} { 

\begin{tabular}{lccc}

\hline
& {\bf S12572-015P} & {\bf S12572-025P} & {\bf S13360-3050P} \\
\hline

Pixel Size ($\mu$m) &  15     &    25   &  50  \\
Fill Factor &    0.53     &    0.65	 &  0.74  \\
Active Area (mm$^2$)  &  4.77   &   5.85  &  6.66   \\
Operating Voltage ($V_{bd}$+V) & +4V & +3.5V & +3.0V \\
Gain at $V_{op}$  &  2.35 $\times 10^5$   &  5.15 $\times 10^5$   &   1.70 $\times 10^6$  \\

\hline
& \multicolumn{3}{c} {\bf Currents (nA/Gain/mm$^2$)}  \\
\hline

{\bf Gammas} & & & \\
\multicolumn{1}{c} {Initial}  & 3.62 $\times 10^{-5}$  &   2.62 $\times 10^{-5}$   &   1.16 $\times 10^{-5}$ \\
\multicolumn{1}{c} {1 Mrad}  & 1.53  $\times 10^{-3}$  &  1.04 $\times 10^{-3}$  &   9.08 $\times 10^{-4}$ \\
\multicolumn{1}{c} {Ratio} & 42.3 &  39.9  &  78.5 \\
{\bf Neutrons} & & & \\
\multicolumn{1}{c} {Initial} & 3.41 $\times 10^{-5}$  &   2.96 $\times 10^{-5}$   &   1.21 $\times 10^{-5}$ \\
\multicolumn{1}{c} {$10^{10}$ n/cm$^2$} &  3.11 $\times 10^{-2}$   &   3.39 $\times 10^{-2}$   &   2.56 $\times 10^{-2}$ \\
\multicolumn{1}{c} {Ratio} &  912.5  & 1144.4  &  2108.3  \\
{\bf Neutrons$/$Gammas} & & & \\
\multicolumn{1}{c} {$1.8 \times 10^{10}$ n/cm$^2$ / 1 Mrad} &  20.3  & 32.5  &  28.2  \\

\hline
\end{tabular} 
}
\label{tab:radiation_induced_currents}
\end{table}

We also investigated the effect of radiation damage on the window material used for the Hamamatsu MPPCs. The main concern for radiation damage in the window is that radiation can produce color centers that can lead to optical absorption, thus leading to a loss in the effective Photon Detection Efficiency (PDE) of the device. 

We measured the optical transmission of several types of window materials before and after an exposure to 1 Mrad of gamma ray dose. Four types of window materials were tested: epoxy resin and silicone (which has a transmission wavelength cutoff deeper in the UV), both of which came in two forms: potting and molding. The S12572s tested in our other irradiation studies were supplied with a molded epoxy window. Samples of the window materials were prepared with a 0.30 mm thick layer of the window material sandwiched between two thin fused silica windows to facilitate handling and optical transmission measurements.

\begin{figure}[th]
 \begin{center}
   \includegraphics[width=9cm]{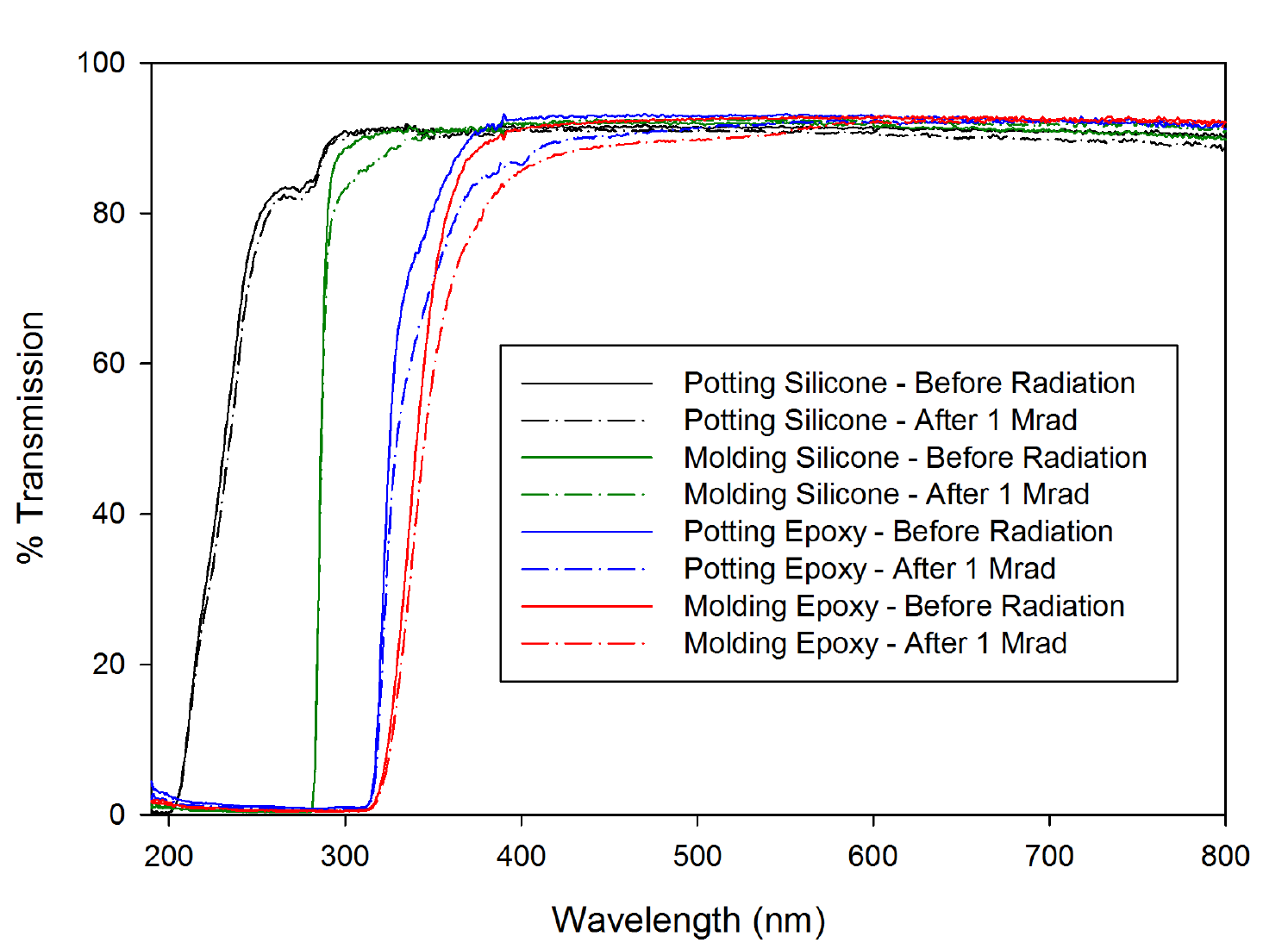}
   	\caption{\label{fig:Gamma_irradiation_window_absorption} Effect on the optical transmission of various types of window materials used in Hamamatsu MPPCs for a cumulative exposure of 1 Mrad of $^{60}$Co gamma rays.}
 \end{center}
\end{figure}

Figure~\ref{fig:Gamma_irradiation_window_absorption} shows the effect of the gamma irradiation on various window materials. Potting silicone showed the least amount of damage, while the molded silicone showed slightly more damage near the band edge. Both the potting epoxy and molding epoxy showed more damage, but in absolute terms, the damage was not severe, even at the highest dose.  The transmission loss was only $\sim$ 5-7\% for wavelengths greater than 400 nm, which is the region for emission for many common scintillators. 

\subsection{Neutron Irradiations}

\begin{figure}[th]
 \begin{center}
   \includegraphics[width=9cm]{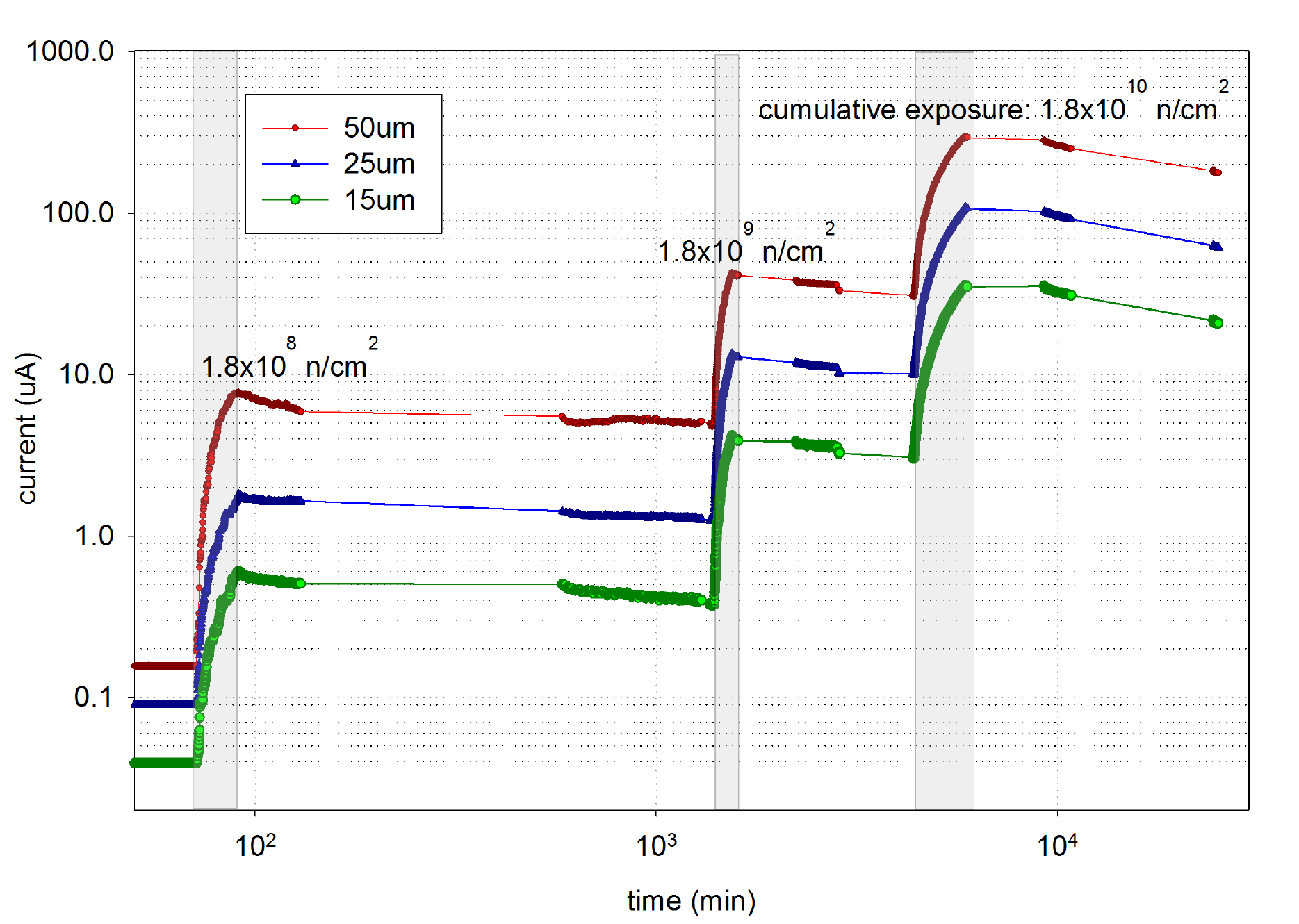}
   	\caption{\label{fig:Neutron_irradiation_SSIF} Sequence of irradiations for Hamamatsu 15 $\mu$m, 25 $\mu$m and 50 $\mu$m pixel devices up to $1.8 \times 10^{10}$ n/cm$^2$ at the BNL SSGRIF  at a flux of 10$^5$ n/cm$^2$/sec. Currents were measured at the manufacturers recommended operating voltage for each device listed in Table~\ref{tab:radiation_induced_currents}.}
 \end{center}
\end{figure}

Neutron irradiations at the SSGRIF were carried out in a series of steps, similar to those done with gamma rays. The flux was fixed throughout the exposures at a rate of 10$^{5}$ n/cm$^2$/sec. Figure~\ref{fig:Neutron_irradiation_SSIF} shows the results for exposures of $1.8 \times 10^8$, $1.8 \times 10^9$ and $1.8 \times 10^{10}$ n/cm$^2$ for the 15 $\mu$m, 25 $\mu$m and 50 $\mu$ pixel devices. In all cases, the dark current increases steadily during the irradiation and then decreases slowly after the source is removed, in contrast to the behavior seen with the gamma ray exposures. The slow decrease after exposure to neutrons is evidence for room temperature annealing, although the amount of recovery is rather small. 

A similar set of neutron exposures were also carried out at the Atomki cyclotron where higher fluxes could be achieved. Exposures were made in various steps at 3.5$\times 10^5$ n/cm$^2$/sec, 1.5$\times 10^6$ n/cm$^2$/sec, 8.0$\times 10^6$ n/cm$^2$/sec and 7.0$\times 10^7$ n/cm$^2$/sec, which made it  possible to reach a maximum fluence of 10$^{12}$ n/cm$^2$. 
 
\begin{figure}[th]
 \begin{center}
	\includegraphics[width=9cm]{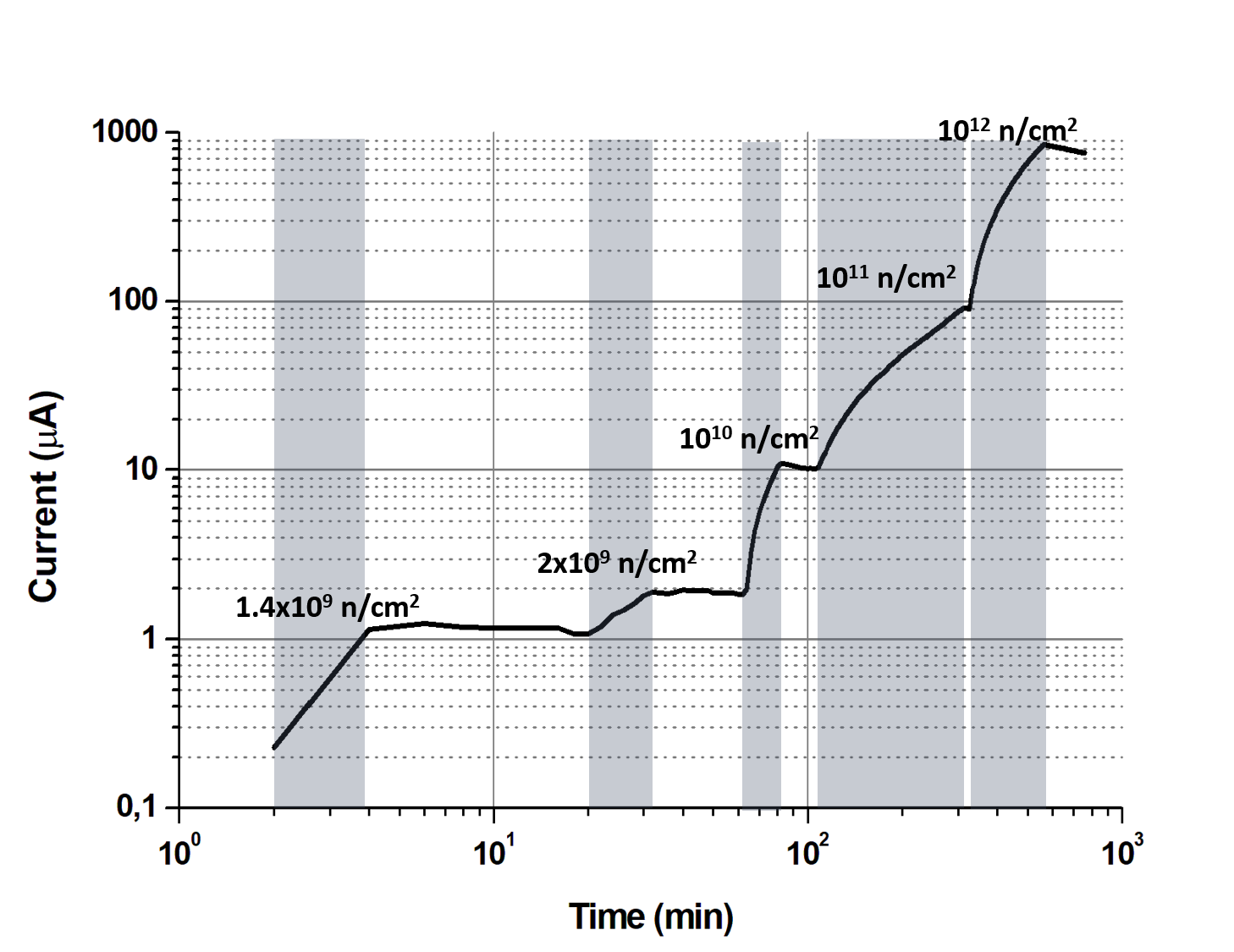}
\caption{\label{fig:Neutron_irradiation_Atomki} Sequence of irradiations  of a Hamamatsu S12572-015P with neutrons at the Atomki cyclotron reaching a maximum fluence of $10^{12}$ n/cm$^2$. Currents were measured at the manufacturers recommended operating voltage listed in Table~\ref{tab:radiation_induced_currents}.}
 \end{center}
\end{figure}

 Figure~\ref{fig:Neutron_irradiation_Atomki} shows the increase in dark current for the exposures of a S12572-015P in five steps from 1.4$\times 10^9$ n/cm$^2$ up to 10$^{12}$ n/cm$^2$. The overall behavior is very similar to what was observed at the BNL SSGRIF. In both cases, there is a steady increase in current during the irradiation, followed by a slow and modest decrease in current after the source is removed. The total dark current reached was $\sim$ 1 mA for the maximum exposure of $10^{12}$  n/cm$^2$. For a fluence of $1.8 \times 10^{10}$ n/cm$^2$ at the SSGRIF, the current reached a value of 35 $\mu$A, while for the Atomki sample irradiated to $1.0 \times 10^{10}$ n/cm$^2$, the current reached was $\sim$ 10 $\mu$A, which would correspond to a current of $\sim$ 18 $\mu$A at the same equivalent fluence as measured at the SSGRIF. We believe this difference of roughly a factor of two is due to the systematic uncertainties in determining neutron fluence at the two facilities, and is primarily determined by the uncertainty in knowing the absolute flux of neutrons for the D-T generator at the SSGRIF.

\begin{figure}[th]
 \begin{center}
   \includegraphics[width=9cm]{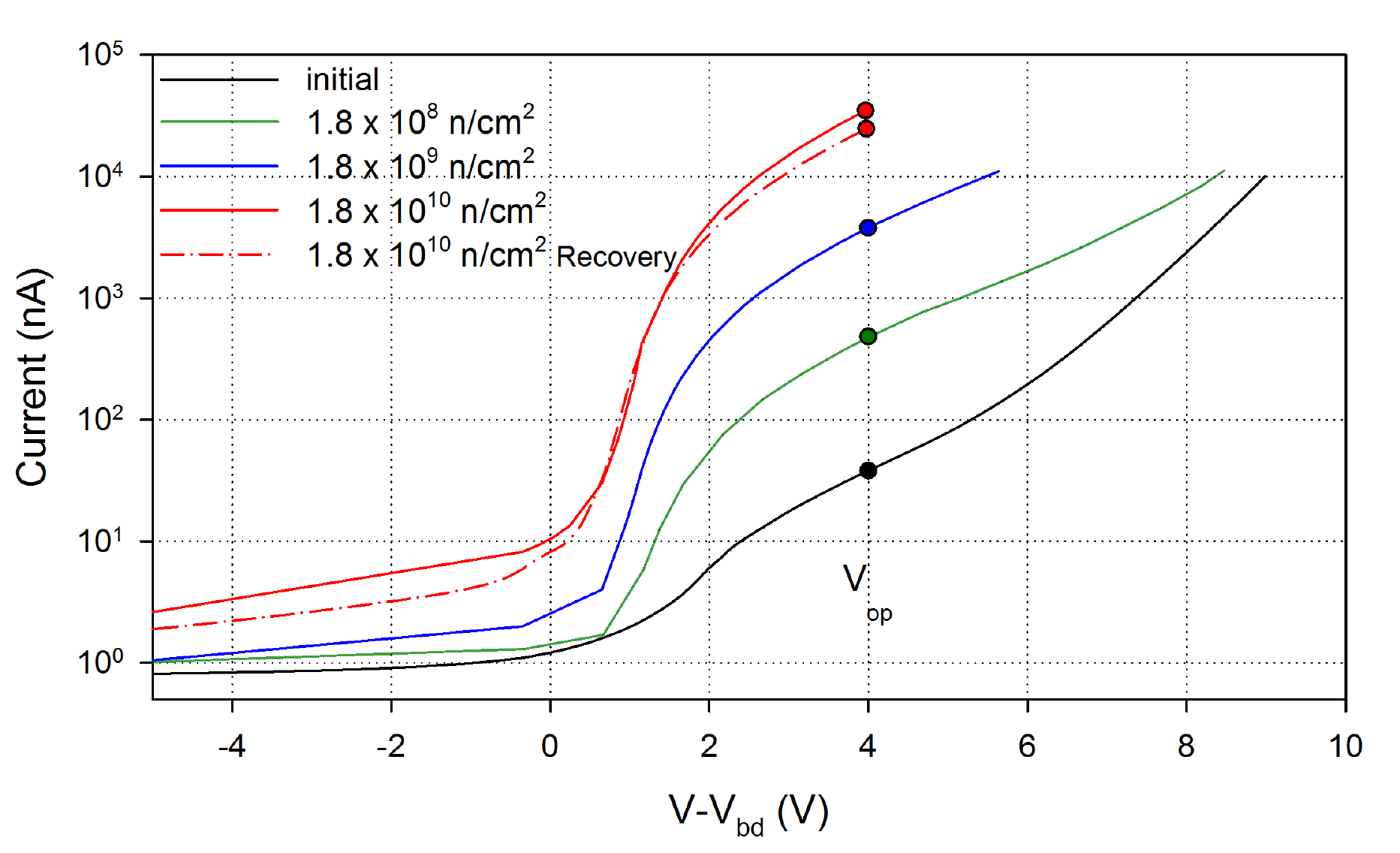}
   	\caption{\label{fig:Neutron_irradiation_I-V_curve} I-V curve for a 15 $\mu$m pixel Hamamatsu S12572-015P irradiated up to $1.8 \times 10^{10}$ n/cm$^2$ at the BNL SSGRIF followed by 5 days of recovery at room temperature.}
 \end{center}
\end{figure}

   Figure \ref{fig:Neutron_irradiation_I-V_curve} shows the I-V curves for a 15 $\mu$m Hamamatsu S12572-015P irradiated at the SSGRIF in steps up to a maximum fluence of $1.8 \times 10^{10}$ n/cm$^2$ along with the I-V curve after 5 days of recovery at room temperature after the maximum exposure. The current reached a maximum of $\sim$ 35 $\mu$A, similar to what was observed in Fig. 
\ref{fig:Neutron_irradiation_SSIF}, and showed very little recovery. However, the largest increase in current occurred only above the breakdown voltage, in contrast to the gamma ray irradiations, indicating that the damage occurred more in the bulk material as opposed to at the surface.

\begin{figure}[th]
 \begin{center}
   \includegraphics[width=9cm]{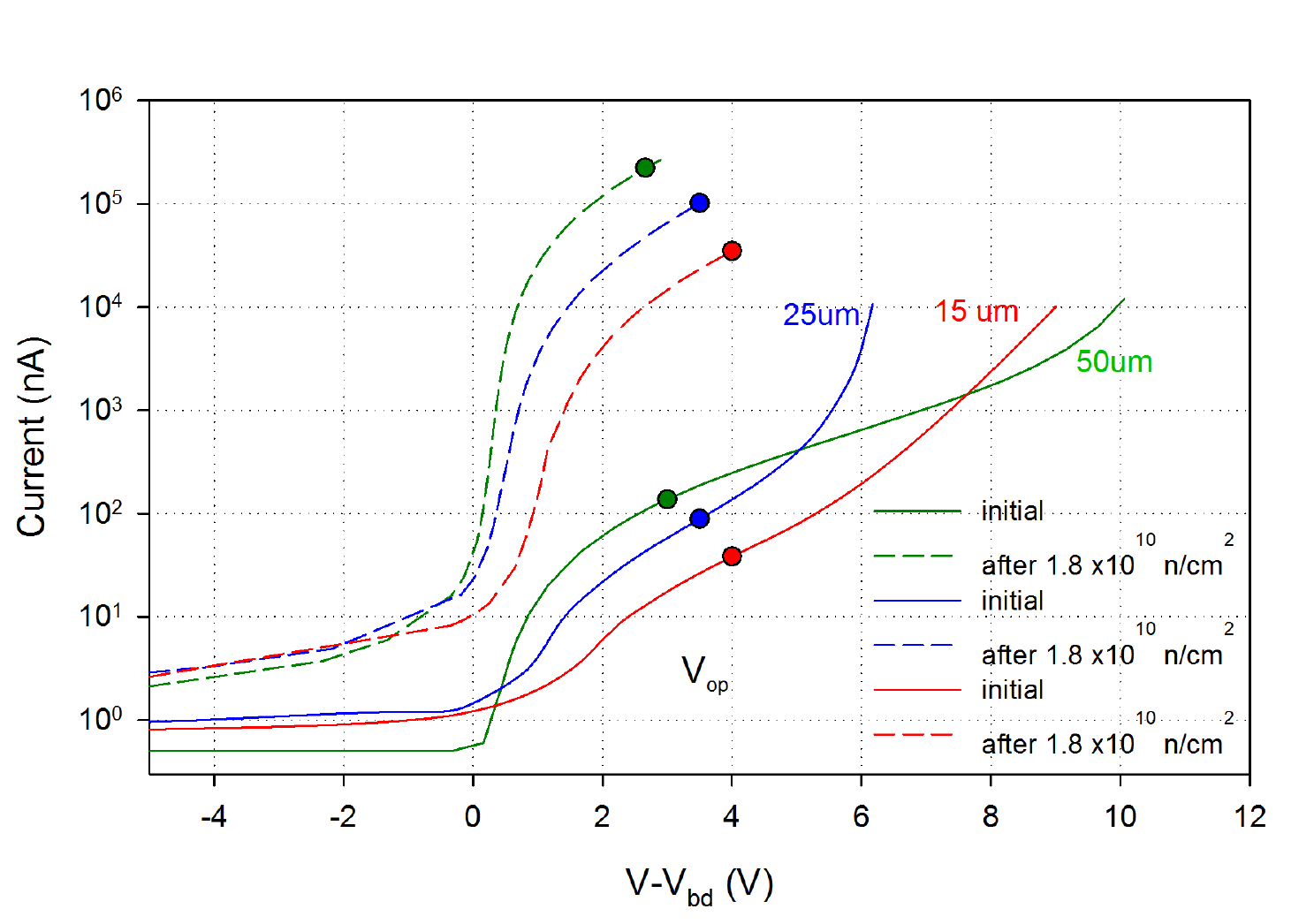}
   	\caption{\label{fig:Neutron_irradiation_3pixel_I-V_curve} I-V curves for 15 $\mu$m, 25 $\mu$m and 50 $\mu$m pixel devices irradiated up to $1.8 \times 10^{10}$ n/cm$^2$ at the BNL SSGRIF. }
 \end{center}
\end{figure}

Figure \ref{fig:Neutron_irradiation_3pixel_I-V_curve} shows the I-V curves for three devices with different pixel sizes that were irradiated up to a maximum fluence of $1.8 \times 10^{10}$ n/cm$^2$ at the SSGRIF. A comparison of the normalized radiation induced currents for the three devices is given in Table~\ref{tab:radiation_induced_currents} along with the similar comparison for the gamma ray exposures. The factors of increase for each of the devices is much larger for a fluence of 1.8x$10^{10}$ n/cm$^2$ compared to the increase factors for 1 Mrad of gamma radiation. The largest increase factor is for the S13360, which may again be due to the fact that its initial dark current was lower. However, if one compares the ratio of the neutron induced current at a fluence of 1.8x$10^{10}$ n/cm$^2$ to the gamma ray induced current at a dose of 1 Mrad, one finds that this ratio is fairly constant for all three devices (in the range $\sim$ 20-30) as shown in the table.   

\begin{figure}[th]
\begin{center}
   \includegraphics[width=9cm]{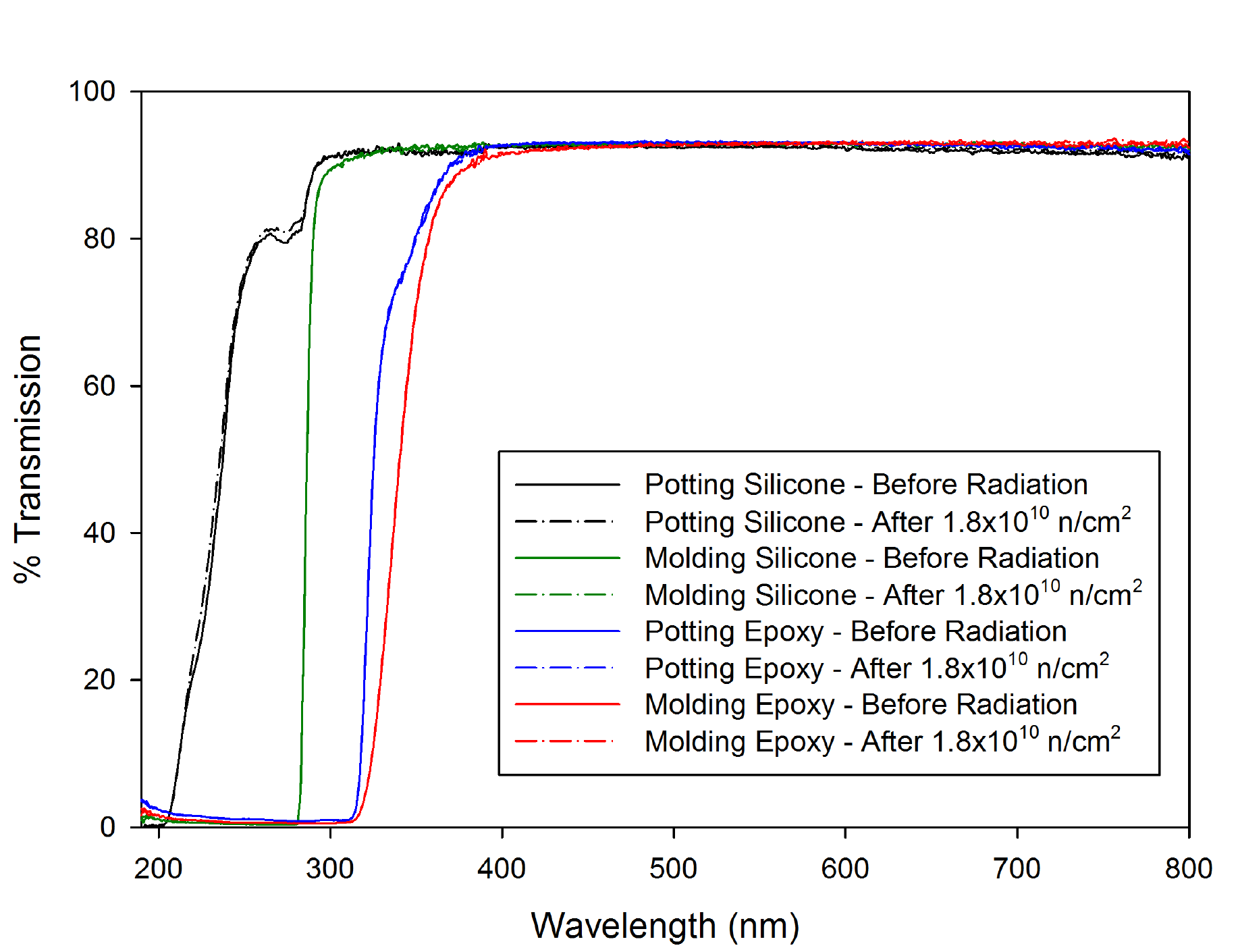}
   	\caption{\label{fig:Neutron_irradiation_window_absorption} Effect on the optical transmission of various types of window materials used in Hamamatsu MPPCs for a total fluence of $1.8 \times 10^{10}$ n/cm$^2$.}
\end{center}
\end{figure}

        The effect of neutron damage to the window materials was also measured. Figure~\ref{fig:Neutron_irradiation_window_absorption} shows the change in absorption of the four types of window materials for a total neutron fluence of $1.8 \times 10^{10}$ n/cm$^2$. No change in absorption was observed for any of the window materials for this level of exposure.

\section{Discussion}

 As discussed in the Introduction, radiation damage in SiPMs can be caused by various types of particles, and this damage can manifest itself in different ways. $^{60}$Co gamma rays ($E_{\gamma} \sim$ 1 MeV) can produce damage by the production of point defects as well as ionization energy loss, the latter occurring through the interaction of the gamma with a nucleus and producing a secondary electron, primarily via Compton scattering. However, due to the fact that NIEL for electrons is significantly lower than for heavier particles, we do not expect that significant displacement damage would be produced in the devices that were tested for the levels of exposure that were attained (see Ref.\cite{Summers:1993} for a comparison of the equivalent electron fluence for 1 Mrad of $^{60}$Co gamma rays). We therefore expect that most of the damage produced by gamma rays for these irradiations is due to ionization energy loss.
  
    A large increase in dark current was observed during irradiation, which is the result of directly producing charge carriers in the silicon due to ionization. However, this current disappeared after the radiation source was removed, but the background dark current slowly increased with dose. This current increased below the breakdown voltage as well as above the breakdown voltage. The large increase below the breakdown voltage indicates a contribution from surface effects due to ionization that does not involve avalanching which occurs deeper inside the device. The background current reached the level of a few $\mu$A for a dose of 1 Mrad and saturated at high doses, and showed very little annealing after irradiation. 

The damage produced by neutrons was consistent with what is expected for bulk damage in the silicon. The main effect was to cause an increase in dark current above the breakdown voltage, implying that this was produced by charge reaching the avalanche region deep inside the device. There was no sudden increase in dark current at the start of the irradiation, as was the case for ionizing radiation, but the background current increased slowly and steadily with exposure. The radiation induced current reached a level $\sim$ 1 mA for a 15 $\mu$m pixel device for an exposure of 10$^{12}$ n/cm$^2$. The dark current remained high after the source was removed and exhibited some annealing at room temperature over a period of minutes to days. There was very little increase below the breakdown voltage, but a large increase above the breakdown voltage, which is consistent with what one would expected for displacement damage in the bulk material. 

 Devices with different pixel sizes behaved similarly for both neutrons and gammas when normalized to their induced current per unit active area and correcting for differences in gain. The ratio of the normalized currents for a neutron fluence of $1.8 \times 10^{10}$ n/cm$^2$ to a gamma ray dose of 1 Mrad was also similar in all devices. 

The optical window materials tested showed some effect due to gamma radiation in the form of reduced optical transmission near the absorption edge. However, the effect was rather small, resulting in a $\sim$  5-7$\%$ transmission loss above 400 nm for 1 Mrad of dose. The induced absorption in silicon window material was concentrated near the knee of the band edge, while the induced absorption in epoxy was produced over a broader range of wavelengths. However, the absorption effects in either of these materials would not pose a problem in terms of significantly reducing the PDE for the levels of exposures tested. There were no significant effects in either the silicone or epoxy window materials for neutron exposures up to $1.8 \times 10^{10}$ n/cm$^2$.

  In addition to measuring the radiation induced currents, we also measured the breakdown voltage in some of the devices after each step of the irradiation. At Atomki, these measurements were done in situ by measuring the I-V curves and computing the breakdown voltage using the method described in Ref.\cite{Nagy:2017}. It was found that there was no change in breakdown voltage up to a maximum dose of $3 \times 10^{11}$ n/cm$^2$ as long as the devices were not held at their operating voltage or higher for any significant time (i.e., only for the I-V scans). However, if the devices were held at their operating voltage or higher for more than a few minutes, then the breakdown voltage was found to increase. If, after performing the I-V scans with a high standing current, the voltage was reduced and the scans performed with a low standing current, no change in breakdown voltage was observed. This suggests that the device heats up due to the high standing current, thereby increasing the junction temperature and causing a change in the breakdown voltage. We are currently pursuing further tests to confirm this hypothesis and will report on this at a future time.
  
   The main purpose of this study was to investigate the performance of various types of SiPMs for use in nuclear physics experiments where the levels of exposure are moderate ($\sim$ 10$^{10}$ n/cm$^2$ and a few 10's of krad per year at RHIC and probably much less at EIC), which are much less than those expected at LHC. The sPHENIX experiment at RHIC will use more than 100K SiPMs for its electromagnetic and hadronic calorimeters and will use the Hamamatsu S12572 15 $\mu$m pixel sensor. This device, which has 40K pixels, was selected in order to have a large dynamic range and reasonably high PDE (25\%). sPHENIX is being designed to cope with the expected increase in noise and dark current by providing sufficient power to deliver up to 250 $\mu$A per device and sufficient cooling to maintain an operating temperature as low as 0 $^{\circ}$C if required.   


\section{Conclusions}

We have carried out a study of radiation damage in various types of Hamamatsu 3 x 3 mm$^2$ MPPCs produced by gammas and neutrons in order to evaluate their performance when exposed to moderate levels of neutrons and gamma rays. The effects of the damage caused by gammas and neutrons manifest themselves differently in these devices due to the differences in their damage mechanism. In both cases, the main effect of the damage is to produce a large increase in the dark current which persists long after the radiation source is removed with only slight recovery at room temperature. The radiation induced dark current caused by gammas has a significant contribution below the breakdown voltage, while neutrons produce a much larger increase in dark current above the breakdown voltage. The result is consistent gammas producing mainly damage due to ionization while neutrons produce mainly displacement damage in the bulk material. Little or no effect was observed in the optical transmission of the SiPM window materials by either neutrons or gammas. Finally, these studies were carried out in order to investigate the radiation effects on various types of SiPMs for use in nuclear physics experiments and enabled the sPHENIX experiment to design its calorimeter system for the expected radiation levels at RHIC. 

\section*{Acknowledgments}

We want to thank the technical staffs of Brookhaven National Laboratory and the Atomki Institute for Nuclear Research for assistance in carrrying out these measurements. This work was supported in part by the U.S. Department of Energy under Prime Contract No. DE-SC0012704 and by the VKSZ\_14-1-2015-0021 project financed from the National Research, Development and Innovation Fund of Hungary in the framework of the Sz\'echenyi 2020 program.

\thanks{G. David is with Stony Brook University, Stony Brook, NY and Brookhaven National Laboratory, Upton, NY 11973-5000} 

\thanks{J.S. Haggerty, E.J. Mannel, S. Stoll,
 and C.L. Woody are with Brookhaven National Laboratory, Upton, NY 11973-5000.}

\thanks{T. Majoros and B. Ujvari are with the University of Debrecen, Debrecen, Hungary.}

\thanks{B. Bir\'o, A. Fenyvesi, J. Moln\'ar and F. Nagy are with Institute for Nuclear Research, Hungarian Academy of Sciences (Atomki), 18/c. Bem t\'er, H-4026, Debrecen, Hungary}



\bibliographystyle{IEEEtran}
\bibliography{IEEEabrv,bibfile}

\end{document}

%% file: defs.tex
\usepackage{xspace}

%% file: introduction.tex
\label{sec:introduction}